\documentclass[12pt]{article}
\usepackage{times}
\usepackage{geometry}
\usepackage[utf8]{inputenc}
\usepackage{enumitem,amssymb}
\usepackage{ragged2e}
\newlist{thematic}{itemize}{8}
\setlist[thematic]{label=$\square$}
\usepackage{pifont}
\newcommand{\cmark}{\ding{51}}%
\newcommand{\done}{\rlap{$\square$}{\raisebox{2pt}{\large\hspace{1pt}\cmark}}%
\hspace{-2.5pt}}

\usepackage{graphicx}
\usepackage{url}
\usepackage{amsmath}
\usepackage{jheppub}   
\usepackage[dvipsnames,svgnames,x11names]{xcolor}
\usepackage[outercaption]{sidecap}
\sidecaptionvpos{figure}{c}
\newcommand{\arcsec}{\hbox{$^{\prime\prime}$}}

\begin{document}

\raggedright
\huge
Astro2020 Science White Paper \linebreak

The warm and dense Galaxy - tracing the formation of dense cloud structures out to the Galactic Center\linebreak
\normalsize

\noindent \textbf{Thematic Areas:} \hspace*{60pt} $\square$ Planetary Systems \hspace*{10pt} \done Star and Planet Formation \hspace*{20pt}\linebreak
$\square$ Formation and Evolution of Compact Objects \hspace*{31pt} $\square$ Cosmology and Fundamental Physics \linebreak
  $\square$  Stars and Stellar Evolution \hspace*{1pt} $\square$ Resolved Stellar Populations and their Environments \hspace*{40pt} \linebreak
  $\square$    Galaxy Evolution   \hspace*{45pt} $\square$             Multi-Messenger Astronomy and Astrophysics \hspace*{65pt} \linebreak
  
\textbf{Principal Author:}

Name: Thomas Stanke
 \linebreak						
Institution: ESO 
 \linebreak
Email: tstanke@eso.org
 \linebreak
Phone:  +49 89 3200 6116
 \linebreak
 
\textbf{Co-authors:} Henrik Beuther (MPIA), Jens Kauffmann (MIT Haystack), Pamela Klaassen (STFC), Juan-Pablo P\'{e}rez-Beaupuits (ESO), Doug Johnstone (NRC-Herzberg), Dario Colombo (MPIfR Bonn), Alvaro Hacar (Leiden), Frederic Schuller (MPIfR Bonn), Sarah Sadavoy (CfA), Juan Soler (MPIA Heidelberg), Jennifer Hatchell (Exeter), Stuart Lumsden (Leeds), Craig Kulesa (Steward Observatory)
  \linebreak

\textbf{Abstract: The past two decades have seen extensive surveys of the far-infrared to submillimeter continuum emission in the plane of our
Galaxy. We line out prospects for the coming decade for corresponding molecular and atomic line surveys which are needed to fully understand the formation of the
dense structures that give birth to clusters and stars out of the diffuse interstellar medium. We propose to work towards Galaxy wide
surveys in mid-J CO lines to trace shocks from colliding clouds, Galaxy-wide surveys for atomic Carbon lines in order to get a detailed
understanding of the relation of atomic and molecular gas in clouds, and to perform extensive surveys of the structure of the dense parts of
molecular clouds to understand the importance of filaments/fibers over the full range of Galactic environments and to study how dense cloud
cores are formed from the filaments. This work will require a large (50\,m) Single Dish submillimeter telescope equipped with massively multipixel
spectrometer arrays, such as envisaged by the AtLAST project.}

\pagebreak
\justify

\section{Introduction}

The last 10-15 years have seen the advent of comprehensive continuum surveys
of the Galactic plane, covering large portions of the plane, at wavelengths
ranging from the thermal infrared (Spitzer) over the far-infrared (Herschel) to
the submillimeter and millimeter regime (e.g., GLIMPSE \cite{benjaminetal2003}, MIPSGAL \cite{careyetal2009}, HIGAL \cite{molinarietal2010}, ATLASGAL \cite{schulleretal2009}, Bolocam Galactic Plane
survey \cite{aguirreetal2011}). The spatial resolution of these surveys ranges from a few arcseconds
(Spitzer) to a few ten arcseconds (Bolocam, Herschel).

\begin{SCfigure}[10.5][h!]
  \hspace{-1mm}
   \includegraphics[width=80mm]{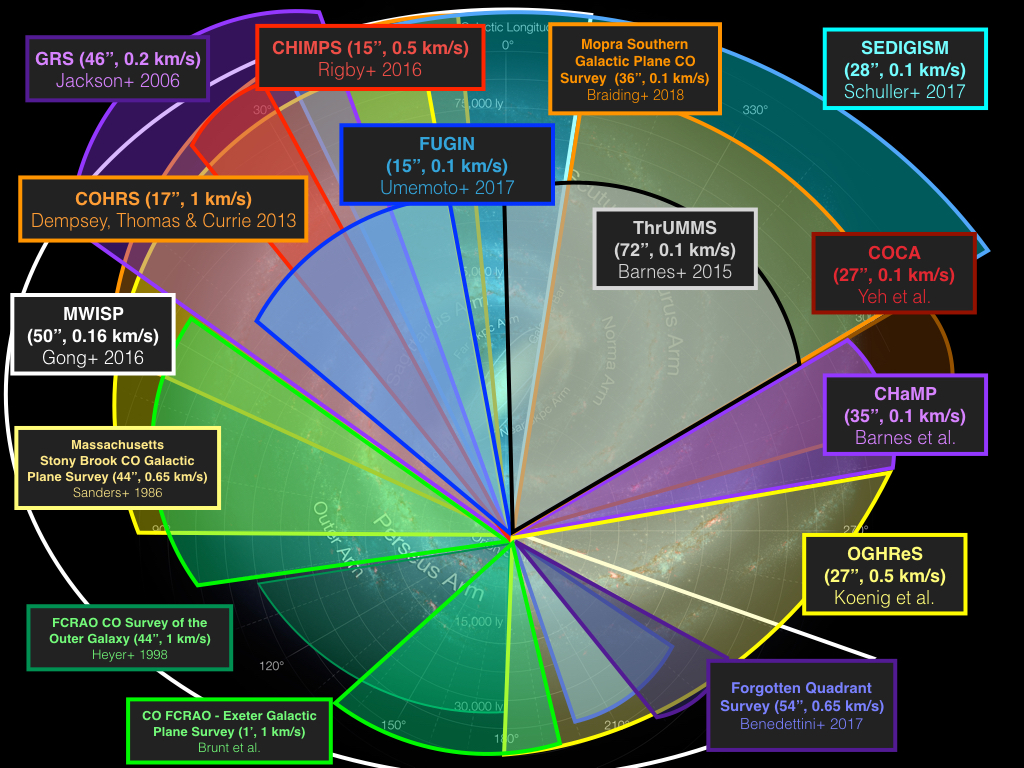}
   \caption[This is a caption]{\small Existing Galactic Plane CO surveys\newline
   \hspace*{-4mm}\begin{tabular}{l|l|l}
    Mopra SGP \cite{braidingetal2018} & CO/$^{13}$O/C$^{18}$O/C$^{17}$O(1-0) & $|b| \leq 0.5$ \\
    SEDIGISM \cite{schuller2017}   & $^{13}$CO/C$^{18}$O(2-1) & $|b| \leq 0.5$ \\
    ThrUMMS \cite{barnesetal2015}  & CO/$^{13}$CO/C$^{18}$O/CN(1-0) & $|b| \leq 1$ \\ 
    Forg. Qu. S. \cite{benedettinietal2017} & CO/$^{13}$CO(1-0) & $|b| = $0..2 \\
    FCRAO Out Gal \cite{heyeretal1998} & CO(1-0) & $b$=-3..5.4 \\
    MassSB\,COGPS \cite{sandersetal1986} & CO(1-0) & $|b| \leq 1$ \\
    MWISP \cite{gongetal2016} & CO/$^{13}$CO/C$^{18}$O(1-0) & $|b| \leq 5.2$ \\
    COHRS \cite{dempseyetal2013} & CO(3-2) & $|b| \leq 0.5$ \\
    GRS \cite{jackson2006} & $^{13}$CO(1-0) & $|b| \leq 1$ \\
    CHIMPS \cite{rigbyetal2016} & $^{13}$CO/C$^{18}$O(3-2) & $|b|  \leq 0.5$ \\
    FUGIN \cite{umemotoetal2017} & CO/$^{13}$O/C$^{18}$O(1-0) & $|b| \leq 1$ \\
    \multicolumn{3}{l}{COCA, CHaMP, OGHReS, CO FCRAO Ex.: in prep}
   \end{tabular}
   }\label{fig:MW_CO_surveys}
\end{SCfigure}

While spectral line observations can provide us with important physical gas properties (temperature, density,
ionisation state, chemistry, kinematics), the coverage in lines is much more patchy, both in
terms of area coverage and spectral coverage. Area covering surveys
only exist for the lowest-J CO lines (1$\to$0 and 2$\to$1) in $^{12}$CO, $^{13}$CO and C$^{18}$O at comparatively low angular
resolution (Fig.\ \ref{fig:MW_CO_surveys}; 20-30\,arcseconds at the best, e.g., \cite{jackson2006,barnesetal2015,schuller2017,umemotoetal2017}), with the only truly Galaxy-wide survey of Dame et al.\ 2001 \cite{dame2001}
providing a spatial resolution of only 8.5\,arcminutes in CO(1$\to$0). The advent of multipixel heterodyne receivers at (sub)millimeter wavelength
telescopes is starting to change the situation (e.g., HARP/JCMT (\cite{dempseyetal2013,rigbyetal2016}), SuperCAM/APEX \cite{kulesaetal2019}; LASMA/APEX (as of 2019)),
but still surveys remain limited to CO and its isotopologues.

The interstellar medium (ISM) is highly structured. Molecular gas is found in
clouds spanning a wide range in mass (few M$_{\odot}$ to few 10$^6$\,M$_\odot$)
and size (few to few 10\,pc). 
Filamentary structures are a common feature of the structure of molecular
clouds and thought to play an important role in the formation of the dense
structures from which stars and clusters form.
Molecular "clumps", with sizes on the order of 1\,pc and densities on the
order of 10$^3$ to 10$^4$\,cm$^{-3}$, may give birth to stellar clusters.  
Dense cloud "cores", with sizes on the order of 0.1\,pc and densities on the
order of 10$^5$ to 10$^6$\,cm$^{-3}$, give birth to
individual stars or small-n multiple systems.

A proper understanding of
the star formation process implies an understanding of how these
structures form, how they evolve, and how varying ambient conditions
influence their properties and evolution.
The location of a molecular cloud within the Galaxy has an impact on
the conditions under which dense structures form: close to the Galactic Center
clouds are highly turbulent and have higher temperature (60-100\,K, e.g., \cite{ginsburgetal2016}), while towards the
outer Galaxy they are much more quiescent and colder; reduced metallicity 
may further influence their structure and chemistry. 
HII regions or supernova remnants will also impact the evolution
of the cloud and its internal structures.

Following the formation and evolution of dense structures, on scales of
0.1-0.2\,pc, over the full range of Galactic environments, i.e., out to distances
on the order of 8\,kpc, will require a spatial resolution of the order of a few (2.5-5) arcseconds.
Wide-field surveys (Galaxy-wide for the more abundant tracers, deeper and more restricted
for less abundant tracers or high-density tracers) are needed but will require significant improvements
in observing capabilities, in particular large format Heterodyne receiver arrays and a large
Single Dish submillimeter telescope such as AtLAST.

\section{Survey projects for the coming Decade}

\subsection{A Galaxy-wide survey for warm molecular gas and shocks}

The temperature and density structure of molecular clouds is determined
by their formation process.
Of relevance here may be the picture of dense structure formation out of
``colliding flows'' in the interstellar medium (ISM). The turbulent nature of
the ISM will
unavoidably result in regions where large scale, moving gas parcels meet
and collide. At the collision front a sheet of denser gas will form, which
is highly unstable to fragmentation, first into filamentary structures,
which then fragment to form clumpy structures within them (e.g., \cite{inoueinutsuka2009,gongostriker2011,gongostriker2015}).
The zones where the incoming flows collide with the dense sheet can be
expected to show up as warm, shocked regions, featuring emission from warm
molecular gas (CO) and/or shock tracers (e.g., \cite{pon2012,pon2015,pon2016b,louvetetal2016}).

CO is the most commonly observed tracer of molecular gas, being abundant
and bright, tracing mostly moderately dense gas ($10^3$ cm$^{-3}$). It is a good
tracer of the bulk of a molecular cloud, although a significant fraction of
molecular gas might be ``CO-dark''(e.g., \cite{ackermannetal2012}). With the main isotopologue becoming optically
thick rapidly, a full account of molecular gas requires covering rarer
isotopologues ($^{13}$CO, C$^{18}$O), and higher energy transitions. The spectral-line energy distribution (SLED)
of CO lines from different rotational levels holds important clues about
the temperature and excitation mechanisms of the CO molecular gas. Covering
intermediate- to high-J CO lines will allow us to trace warm gas, and comparing
line strengths with low-J lines will enable us to discriminate between 
emission from Photon Dominated Regions (PDRs) and low velocity
shocks, e.g., from converging flows forming dense structures \cite{pon2016b}.
Area covering surveys of submillimeter CO lines, at least up to the J=7$\to$6
transition, will be crucial in identifying converging flows and in
quantifying their importance for the formation of clouds and substructures
within them.

\subsection{Cloud formation from the atomic to the molecular phase}

There is still significant debate regarding the transition from the atomic to the  molecular ISM phases.  Are quasi-static contraction processes associated
with ever increasing densities and related conversion of atomic to
molecular gas dominating (e.g., \cite{mckee2007}), or is the
process more dynamic where converging gas flows create over-densities
and within them the gas converts the atomic into the molecular phase
(e.g., \cite{banerjee2009,vazquez2011})?

To investigate these different possibilities, studying the different
gas phases in the molecular as well as atomic environment is
important. While there exist several surveys of the Milky Way in the
molecular phase (Fig.\ \ref{fig:MW_CO_surveys}) as
well as in the neutral atomic hydrogen (e.g.,
\cite{stil2006,winkel2016,beuther2016}), large-scale maps of a good
tracer of the transition phase are still rare and very limited in area coverage
(e.g., \cite{tanaka2011}: [CI] in the Galactic Center; \cite{shimajiri2013}: [CI] in Orion~A). 
One of the best transition phase tracer may be the fine structure lines of carbon [CI] (e.g.,
\cite{glover2015}), but currently the picture is very unclear. [CI] emission at times is found to extend to more
diffuse, CO-dark regions, and thus out to larger size scales,
potentially even the scales at which turbulence may be driven (e.g.,
\cite{brunt2002}). Often, [CI] is also found to be similar in extent to dense material traced, e.g., in C$^{18}$O
(e.g., \cite{perezbeaupuitsetal2015b}; see also \cite{clarketal2019}), with small, but significant differences in
exact location and extent seen in some cases (e.g., \cite{beutheretal2014}).

A large (50\,m) Single Dish submillimeter telescope such as AtLAST would allow the first high-spatial-resolution Galactic plane survey in both atomic carbon transitions at 492\,GHz ([CI]($^3$P$_1-^3$P$_0$, $E_u/k=23.6$\,K) and 809\,GHz ([CI]($^3$P$_2-^3$P$_1$, $E_u/k=62.5$\,K). With a dish size of 50\,m, angular resolution elements at the two transitions of $\sim$3\arcsec{} and $\sim$1.8\arcsec{} will be achieved. At typical molecular cloud distances in the Milky Way of several kpc, that corresponds to linear resolution elements typically significantly below 0.1\,pc;
a careful comparison of these atomic Carbon data with comparable datasets then available for atomic hydrogen from the SKA, and molecular data from AtLAST as well as ALMA will allow us to study the atomic to molecular gas conversion processes as well as cloud formation and destruction mechanisms in great detail.

\subsection{Intermediate scales/Filaments}

Filaments have since long been known to permeate molecular clouds (e.g., \cite{ballyetal1987}). Herschel far-IR
surveys in particular have revealed extensive networks of filaments in
nearby molecular clouds, intimately related to the location of cloud
cores (e.g., \cite{andre2014}). Typically, these filaments have a width
on the order of 0.1-0.2~pc.
More distant filaments (few kpc) such as identified by \cite{lietal2016} from the ATLASGAL
survey are found to cover a much
larger range in widths (0.1-2.5\,pc). Cluster forming massive clumps are
often seen to be located on massive filaments (termed "ridges") and/or at the
intersection of filaments. 

In addition to the continuum, molecular line observations allow to study the internal mass distribution and dynamical properties of filaments.
The analysis of different density selective tracers (i.e., C$^{18}$O and N$_2$H$^+$) in clouds such as Taurus and Orion demonstrates that different nearby filaments are actually collections of the so-called fibers, namely, smaller-scale sub-filaments twisted in space forming a bundle \cite{hacaretal2013,hacaretal2018}. Similar fiber-like arrangements have been also reported in other nearby regions (e.g., \cite{FernandezLopezetal2014}) as well as in more distant Infrared Dark Clouds (IRDCs, e.g., \cite{henshawetal2016}). It remains to be investigated to which extent more massive filaments identified across the Galaxy are indeed monolithic structures, typically unresolved by current single-dish observations, or if they rather separate into sub-filaments when observed at higher spatial resolution. Molecular line studies indicate a widespread detection of multi-scale velocity oscillations and excursions along filaments across their entire mass spectrum (e.g., \cite{jacksonetal2010}). Also, and at sub-parsec scales, sinusoidal velocity fields may indicate gas flows towards the densest portions of fibers, leading to core formation, but have only tentatively been detected (L1517, \cite{hacartafalla2011}). Investigating the nature and magnitude of these motions, only accessible using multi-line surveys, could provide fundamental insights about the formation mechanism of filaments, their fragmentation process, and the origin of stars in them.

\begin{SCfigure}[10.5][h!]
  \hspace{-1mm}
   \includegraphics[width=100mm]{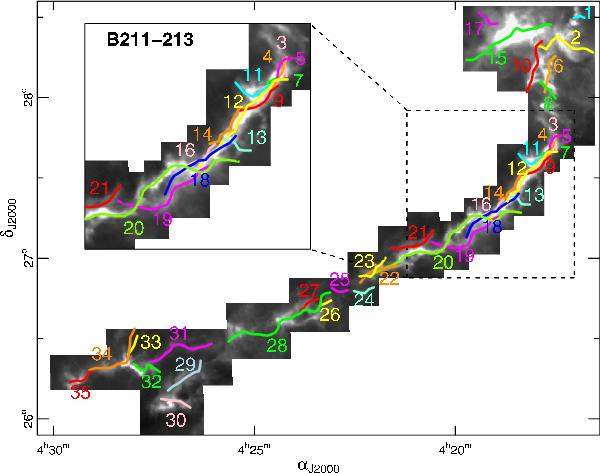}
   \caption[This is a caption]{\small Filament-Fiber structures in Taurus \cite{hacaretal2013}
   }\label{fig:TaurusFilaments}
\end{SCfigure}

Recent molecular maps reported the first evidence of ordered velocity gradients in both clusters \cite{kirketal2013} and IRDCs \cite{perettoetal2014} detected in species like  NH$_3$ and N$_2$H$^+$. 
Converging velocity patters are found along filaments forming hub-like structures, likely funneling material towards the most massive clumps located at the centre of these filamentary associations (e.g., \cite{perettoetal2013}). Global infall velocity patterns detected at parsec scales indicate that young clusters such as the Orion Nebula Cluster continue accreting material after several Myrs \cite{hacaretal2017}. Resolving these motions at high spatial resolutions result key to understand how clusters assemble their masses over time as well as the formation of massive stars within these regions. The study of these motions requires the simultaneous characterization of dense (e.g., low-J lines from density selective species such us N$_2$H$^+$ or HCN) and diffuse tracers (e.g., high-J CO lines) in order to disentangle the strong feedback effects present in these regions. The expected multi-scale, high-dynamic range mapping capabilities of AtLAST offer the first opportunity to systematically study these processes in detail across the Milky Way. 



\subsection{Cores}
Moving on to the cloud cores themselves, our current understanding is still largely based on
detailed studies of individual objects or limited studies (in terms of available tracers) of
moderate sized samples of cores. A full understanding of the physics of core formation and evolution
towards and into star formation will require comprehensive studies of large core samples (comprising
the full range of possible environments) for their physical properties (mass, density and temperature structure,
chemical structure, etc.). E.g., mid-$J$ ($J=7\to6$ to $J=10\to9$) HCN and HCO$^+$ lines available at ALMA Band 9 and 10
(600-950~GHz) trace very dense ($n({\rm H_2})_{cr} > 10^6$~cm$^{-3}$ at $T_K = 100$~K) and warm ($E_u/k > 150$~K) gas
and will allow to identify and study the cores where stars are actually forming (or are about to form; e.g., \cite{perezbeaupuitsetal2015}). 

In the dense cores we shall also see the effects of ambipolar diffusion, expected to be observed in infalling gas in the 
presence of strong magnetic fields as a small difference in the velocity fields between ionized and neutral species 
(e.g., \cite{houdeetal2002,tangetal2018,chingetal2018}). 
Doing this on large samples, covering scales from cluster forming, parsec-sized
clumps down to individual star forming cores in one go, will help to understand better the role of magnetic fields as the star formation process proceeds.
This type of observations requires a combination of decent area coverage (on the order of 10$\times$10~arcmin), spatial resolution
on the order of a few arcseconds, high sensitivity at high spectral resolution, and spectral coverage (multi-line analysis
to rule out optical depth and chemical effects on the line shapes).

\section{The case for a large sub-mm single dish telescope}

To summarize the above, following the formation and evolution of molecular clouds and their structure from
Giant Molecular cloud scales down to cloud cores forming individual stars in a comprehensive manner requires:

\begin{itemize}
    \item Galaxy-wide surveys of tracers of molecular gas (to first order CO isotopologs) and the precursor atomic gas ([CI]);
    \item multiline observations in CO, up to at least mid-J transitions, (and other shock tracers) to identify large scale shocks from cloud collisions;
    \item deep surveys of high density tracers (e.g., HCO$^+$, HCN; covering areas of hundreds of square arcminutes to tens of square degrees) to follow the formation of dense
    structures such as filaments and cores, with angular resolution up to 2-3\,arcsec (cloud core size at the distance of the Galactic Center)
\end{itemize}

These requirements can be met by a 50\,m single-dish telescope operating at submillimeter wavelengths (1\,mm - 250\,$\mu$m), equipped with massively multipixel
($\sim$1000) spectrometers providing wide spectral coverage and high spectral resolution (better than 100\,m/s), as envisaged by the AtLAST project.

\subsection{50\,m Single Dish (AtLAST) vs.\ ALMA:}

ALMA easily provides the required angular resolution, but will not be
competitive in mapping speed (50 12\,m diameter antennas with single-pixel receivers vs.\ one 50\,m antenna with $\sim$1000-pixel receiver array).
In addition, molecular clouds show complex, multiscale emission, 
which will be subject to spatial filtering with ALMA; it becomes more and more
evident that even multi-configuration interferometric observation including
total power observations give a skewed representation of the actual intensity
distribution (e.g., \cite{ossenkopfetal2016}). Single dish observations, covering the required range of angular
scales from very large down to the resolution limit are crucial for a proper recovery of the true source emission
structure particularly in regions having a complex spatial and velocity structure (Hacar et al, in prep).
On the other hand, a 50\,m ATLAST will readily provide data that can be used to
complement ALMA interferometric data as zero-spacings, with a very good overlap in uv-space
with ALMA (shortest ACA~7\,m baselines 10-15\,m); this is also in contrast with CCAT-prime, which (with only 6\,m diameter)
does not overlap with the ALMA interferometric uv-space.

\pagebreak

\def\aj{AJ}%
\def\araa{ARA\&A}%
\def\apj{ApJ}%
\def\apjl{ApJ}%
\def\apjs{ApJS}%
\def\ao{Appl.~Opt.}%
\def\apss{Ap\&SS}%
\def\aap{A\&A}%
\def\aapr{A\&A~Rev.}%
\def\aaps{A\&AS}%
\def\azh{AZh}%
\def\baas{BAAS}%
\def\jrasc{JRASC}%
\def\memras{MmRAS}%
\def\mnras{MNRAS}%
\def\pra{Phys.~Rev.~A}%
\def\prb{Phys.~Rev.~B}%
\def\prc{Phys.~Rev.~C}%
\def\prd{Phys.~Rev.~D}%
\def\pre{Phys.~Rev.~E}%
\def\prl{Phys.~Rev.~Lett.}%
\def\pasp{PASP}%
\def\pasj{PASJ}%
\def\qjras{QJRAS}%
\def\skytel{S\&T}%
\def\solphys{Sol.~Phys.}%
\def\sovast{Soviet~Ast.}%
\def\ssr{Space~Sci.~Rev.}%
\def\zap{ZAp}%
\def\nat{Nature}%
\def\iaucirc{IAU~Circ.}%
\def\aplett{Astrophys.~Lett.}%
\def\apspr{Astrophys.~Space~Phys.~Res.}%
\def\bain{Bull.~Astron.~Inst.~Netherlands}%
\def\fcp{Fund.~Cosmic~Phys.}%
\def\gca{Geochim.~Cosmochim.~Acta}%
\def\grl{Geophys.~Res.~Lett.}%
\def\jcp{J.~Chem.~Phys.}%
\def\jgr{J.~Geophys.~Res.}%
\def\jqsrt{J.~Quant.~Spec.~Radiat.~Transf.}%
\def\memsai{Mem.~Soc.~Astron.~Italiana}%
\def\nphysa{Nucl.~Phys.~A}%
\def\physrep{Phys.~Rep.}%
\def\physscr{Phys.~Scr}%
\def\planss{Planet.~Space~Sci.}%
\def\procspie{Proc.~SPIE}%
\let\astap=\aap
\let\apjlett=\apjl
\let\apjsupp=\apjs
\let\applopt=\ao

\bibliography{bibliography}   
\bibliographystyle{unsrturltrunc6}    

\end{document}